\documentclass[12pt]{article}

\usepackage{latexsym}
\usepackage{graphics}
\newcommand{\be}{\begin{equation}}
\newcommand{\ee}{\end{equation}}
\newcommand{\ben}{\begin{eqnarray}}
\newcommand{\een}{\end{eqnarray}}

\newcommand{\bea}{\begin{eqnarray}}
\newcommand{\eea}{\end{eqnarray}}

\begin{document}

\setlength{\baselineskip}{15pt}
\title{
\normalsize
\mbox{ }\hspace{\fill}
\begin{minipage}{7cm}
UPR-943-T\\
{\tt hep-th/0106286}{\hfill}
\end{minipage}\\[5ex]
{\large\bf  Black hole dynamics from thermodynamics \\in anti-de Sitter space
 \\[1ex]}}
\author{ SangChul Yoon\footnote{bathohms@student.physics.upenn.edu} \\
{\it Department  of Physics
and Astronomy},\\
{\it University of Pennsylvania,
Philadelphia PA 19104-6396, USA}\\
}

\maketitle

\begin{abstract}
We work on the relation between the local thermodynamic
instability and the dynamical instability of large black holes in
four-dimensional anti-de Sitter space proposed by Gubser and
Mitra. We find that all perturbations suppressing the metric
fluctuations at linear order become dynamically unstable when
black holes lose the local thermodynamic stability. We discuss
how dynamical instabilities can be explained by the Second Law of
Thermodynamics.

\end{abstract}

\section{Introduction}
Black holes are very interesting objects from their causal structures in general relativity to
 their quantum mechanical properties. To figure out their physical relevance, we need to
 answer if the complete gravitational collapse of a body results in a black hole rather than
 a naked singularity. The conjecture \cite{Penrose} that nature censors naked singularity was
 proposed in this respect. One of motivation of this is from the fact that black holes in 4-dimensional
 asymptotic Minkowski space are stable: linear perturbations around black hole solutions
 do not give any evolution.

However, it was found in \cite{Gregory1, Gregory2} that black
strings and p-branes are unstable. The basic idea of the
Gregory-Laflamme instability is that whatever has the biggest
entropy is favored. Since a black string has a different topology
of horizon as that of a black hole and entropy is proportional to
the area of horizon, array of black holes has bigger entropy than
a uncompactified black string  when they have the same mass
\cite{Amanda}. The instability of black strings was shown
\cite{Gregory1, Gregory2} by doing perturbation theory. It is a
very interesting question to see what would happen during the
transition between them. It has been argued that violation of
cosmic censorship does occur during this process. Recently it has
been suggested that a black string settles down to a new static
black string solution which is not translationally invariant along
the string \cite{Horowitz}.

Entropy argument used above was revisited  by Gubser and Mitra to
propose that a black brane becomes dynamically unstable when it
is locally thermodynamically unstable \cite{GM1,GM2}. Local
thermodynamic stability is defined as having an entropy which is
concave down as a function of the mass and the conserved charges
\cite{Gubser}. This conjecture was made from the perspective of
AdS/CFT correspondence \cite{Maldacena, Klebanov, Witten}, which
identifies two low energy excitations, both of which are decoupled
from supergravity in flat space, in two low energy descriptions
of superstring theory \cite{Aharony}. Some unstable fluctuation
modes may be excited when there is a thermodynamic instability in
the field theory and according to AdS/CFT the same thing would
happen in AdS \cite{GM2}. A semi-classical proof of above
conjecture using the Euclidean path integral approach to quantum
gravity was given in \cite{Reall}.

The motivation of Gubser-Mitra (GM) conjecture is that Lorentzian
time evolution should proceed so as to increase the entropy. In
this paper, we find that their argument holds for both of the
cases in which the metric fluctuations are suppressed and in
which only the metric fluctuations are turned on. In the former
case, any perturbation for all equal charges of $AdS_4$-$RN$
solutions becomes dynamically unstable when the system loses the
thermodynamic stability and all evolutions increase entropy. In
the latter case, there is no dynamical instability even though
the system is thermodynamically unstable. The stability can be
explained by the fact that entropy would be decreasing if the
perturbation is unstable. We discuss these in section 2 and
section 3. We try to explain dynamical instabilities of black
holes from the Second Law of Thermodynamics in section 4.

\section{Gubser-Mitra analysis and its generalization}
\subsection{$AdS_{4}$-$RN$ black hole}

An electrically charged black hole in the asymptotically $AdS_{4}$
was found in \cite{DuffLiu}. Starting from $\mathcal{N}=8$
supergravity in 4-dimensions, they gauged the rigid SO(8) symmetry
of 28 gauge boson \cite{BdeWit1} and the potential induced from
this gauging makes $AdS_{4}$ a vacuum solution of this theory. The
$AdS_{4}$ black hole solution is made by focusing on $U(1)^4$
Cartan subgroup of SO(8), which is believed to be a consistent
truncation. Only three scalar fields of 70 scalar fields
 in the original theory are kept by working in symmetric gauge. The Lagrangian of this truncated
  theory is,

\bea && \mathcal{L}=\frac{
\sqrt{g}}{2k^2}\left[R-\sum_{i=1}^{3}\left(\frac{1}{2}(\partial
\phi_{i})^2-\frac{2}{L^2}\cosh{\phi_{i}}\right)-
2\sum_{A=1}^{4}e^{\alpha^{i}_{A} \phi_{i}} (F^{(A)}_{\mu
\nu})^2\right] \nonumber\\
\\
&& \mbox{where}  \,\,\,\, \alpha^{i}_{A}=  \left (
\begin{array}{cccc}
 1 & 1 & -1 & -1 \\ 1 & -1 & 1 & -1 \\ 1 & -1 & -1 & 1 \end{array}
\right ). \nonumber \eea The metric signature is $(-+++)$ and
$G_{4}=\frac{1}{4}$. The electrically charged solutions are

\bea ds^2 & =&
-\frac{F}{\sqrt{H}}dt^2+\frac{\sqrt{H}}{F}dz^2+\sqrt{H}z^2d\Omega^2
\nonumber\\
 e^{2\phi_1} &=& \frac{h_1 h_2}{h_3 h_4}\,\,\,\,\,\,
e^{2\phi_2}=\frac{h_1 h_3}{h_2
h_4}\,\,\,\,\,\,e^{2\phi_3}=\frac{h_1
h_4}{h_2 h_3} \nonumber\\
F^{(A)}_{0z} &=&\pm \frac{1}{\sqrt{8}h^{2}_{A}}
\frac{Q_{A}}{z^2}\\
H &=& \prod^4_{A=1}h_{A}\,\,\,\,\,\,
F=1-\frac{\mu}{z}+\frac{z^2}{L^2}H\,\,\,\,\,\,\,
h_{A}=1+\frac{q_{A}}{z}\nonumber\\
Q_{A}&=&\mu \cosh \beta_{A} \sinh \beta_{A} \,\,\,\,\,\, q_{A}=\mu
\sinh^2 \beta_{A}\nonumber
 \eea
where the quantities $Q_{A}$ are the physical conserved charges.
The mass is \cite{CveticGubser1} \be
M=\frac{\mu}{2}+\frac{1}{4}\sum_{A=1}^{4}q_{A}, \ee and the
entropy is \be S= \pi z^2_{H}\sqrt{H(z_{H})} \ee where $z_{H}$ is
the largest root of $F(z_{H})=0$. It is possible to express $M$
directly in terms of the entropy and the physical charges in the
large black hole limit, $M$$\gg$$L$, as \cite{GM2} \be
M=\frac{1}{2\pi^{\frac{3}{2}}L^{2}\sqrt{S}}\left[\prod^{4}_{A=1}(S^2+\pi^{2}L^{2}Q_{A}^{2})\right]^{\frac{1}{4}}.
\ee We are going to study in the case where all four charges are
equal, $q_{A}=q$. In this case the solution can be written in term
of a new radial variable, $r=z+q$, and it becomes

\bea ds^2&=&-fdt^2+\frac{dr^2}{f}+r^2d\Omega^2 \nonumber\\
F_{0r}&=&\frac{Q}{\sqrt{8}r^2}\\
f&=&1-\frac{2M}{r}+\frac{Q^2}{r^2}+\frac{r^2}{L^2}.\nonumber \eea

It is known \cite{BdeWit2} that the consistent $S^7$ truncation of
11-dimensional supergravity is equivalent to $\mathcal{N}=8$
4-dimensional gauged supergravity. Also the equivalence of large
R-charged black holes in D=4, D=5 and D=7 with spinning
near-extreme M2, D3 and M5 branes are
 respectively demonstrated in \cite{CveticGubser1}. It is important to check that our black
 holes
 can be embedded to higher dimensional black objects because GM conjecture requests the non-compact
 translational symmetry. Any instability found in the large black hole limit, $M$$\gg$$L$ in our case
 implies the instability of M2-brane.

\subsection{Thermodynamic instability and adiabatic evolution}
Local thermodynamic stability is defined as having an entropy
which is concave down as a function of the extensive variables.
This means that the Hessian matrix,

\be \boldmath H \unboldmath ^{S}_{M,Q_{A}}\equiv  \left (
\begin{array}{cc} \frac{\partial^2 S}{\partial M^2} &
\frac{\partial^2 S}{\partial M \partial Q_{B}} \\
\frac{\partial^2 S}{\partial Q_{A}\partial M} & \frac{\partial^2
S}{\partial Q_{A}\partial Q_{B}} \end{array} \right )  \ee has no
positive eigenvalues. It is straightforward to express $\boldmath
H \unboldmath ^{S}_{M,Q_{A}}$ in terms of derivatives of
$M(S,Q_{A})$. From this definition, if we introduce the
dimensionless variable $\chi$$ =$$ \frac {Q} {M^{\frac{2}{3}}
L^{\frac{1}{3}}}$ for all equal charges, $Q_{A}$$=$$Q$, the
thermodynamic instability is present when $\chi > 1$ \cite{GM2}.
We can see that the most positive eigenvetor\footnote{The positive
eigenvector here means an eigenvector with a positive
eigenvalue.} of Hessian increases entropy most when entropy is at
its extremum. \be S(M+\delta M,Q_{A}+\delta
Q_{A})=S(M,Q_{A})+\frac{1}{2}(\delta M,\delta Q_{A})\boldmath H
\unboldmath ^{S}_{M,Q_{A}}  {\delta M \choose \delta Q_{B}} \ee
Even though entropy is not at its extremum, we can forget about
the first derivative parts by energy and charge conservation in
microcanonical ensemble. With this it was  found that in the
positive eigenvector direction of Hessian for all equal charges,
the dynamical instability coincides with the thermodynamic
instability with a small discrepancy due to numerical errors. It
will be interesting to see what would happen in other directions.

Gubser and Mitra analyzed the linear perturbation in which fluctuations of the
metric are suppressed. The most unstable eigenvector is $(\delta
M, \delta Q_{A})=(0,1,1,-1,-1)$ for all equal charges. The
condition that the metric decouples at linear order is that
$Q_{A}\cdot \delta Q_{A}=0$. In this case $\delta T_{ab}$
vanishes at linear order and we can also see from (2) the metric
does not change at linear order. It is not difficult to make the
linear perturbation equations in this decoupling case beyond the
eigenvector direction. Our original motivation was to see two
things: first, even though the system loses the thermodynamic
stability, it would not be dynamically unstable if we perturb the
system in the way of decreasing its entropy. Second, because the
eigenvector direction increases entropy most, it would be the
fastest way of increasing entropy.

The general perturbation in which the metric decouples for all
equal charges is that \be \delta Q_{A}=(1,a,b,-a-b-1) \delta Q\ee
where $a$, $b$ can be any real numbers. From (2), we can make an
ansatz about a relevant perturbation \bea \delta
\phi_{i}&=&(1+a,1+b,-a-b)\frac{\delta
\phi}{2} \nonumber\\
\delta F^{(A)}&=&(1,a,b,-a-b-1)\delta F. \eea This ansatz
relating three scalar fields to one scalar field and four U(1)
fields to the other U(1) field should be checked if it is
consistent with equations of motion and it turns out  to be
consistent. The linear perturbation for each $\phi_{i}$ is the
same up to overall factor \be
\left[\nabla_{\mu}\nabla^{\mu}+\frac{2}{L^2}-8F^2_{\mu\nu}\right]\delta
\phi-16F^{\mu\nu}\delta F_{\mu\nu}=0 \ee and the linear
perturbation for each $F^{(A)}_{\mu\nu}$ is also the same up to
overall factor \be d\delta F=0 \,\,\,\,\,\,\,\,\,\,\, d\ast\delta
F+d \delta \phi \wedge \ast F=0. \ee Here $F$ is the background
field strength in (6): it is the same for four $F^{(A)}$. It is
remarkable that all directions have the same perturbation
equation. The case of $a\!=\!1$, $b\!=\!-1$ is the unstable
eigenvector and we can see that (11) and (12) are exactly what
Gubser and Mitra found \cite{GM1}. From their analysis, we can
conclude that all perturbations suppressing metric fluctuation at
linear order have dynamical instabilities when the system is
thermodynamically unstable. We can see that the eigenvector
direction is the fastest way of increasing entropy not because it
evolves fastest but because it increases entropy most. It was
suspected that some of perturbations (9) would decrease entropy
by the continuity of (8) in the case that $\chi$ is slightly
greater than 1, but the Second Law of Thermodynamics is not
violated in another remarkable way\footnote{We thank a referee of
JHEP for pointing out this and we would like to apologize to the
authors of \cite{GM2} for incorrectly disclaiming their argument
in the previous version of this paper.}\footnote{The violation of
the Second Law of Thermodynamics  was suspected from the fact
that AdS is not globally hyperbolic. We thought the area law for
black holes  might not hold. However the area law needs only a
partial Cauchy surface \cite{Hawking}, so it holds in AdS.
$\phi_{i}$ do not satisfy the dominant energy condition, but they
satisfy the strong energy condition defined in \cite{Wald} or the
timelike convergence condition defined in \cite{Hawking}.
Therefore, the area of horizon can not be decreasing with
classical evolutions by the area law.}. The second derivative
parts of (8) with perturbation (9) is \be \frac{1}{2}(\delta
M,\delta Q_{A})\boldmath H \unboldmath ^{S}_{M,Q_{A}}  {\delta M
\choose \delta Q_{B}} =
A(M,Q_{A})(\chi-1)\{1+a^2+b^2+(1+a+b)^2\}.\ee Here $A(M,Q_{A})$
is a positive when $\chi \sim 1 $. We can see that all
perturbations (9) increase entropy when $\chi > 1$. The
factorization of the eigenvalue part $\chi-1$ can explain why we
have the same linear perturbation equations (11) and (12) for all
perturbations (9).

\section{Stability from the metric perturbation analysis}
\subsection{Metric perturbation equation}
In the previous section, we observed that there is a dynamical
instability in any case that the metric perturbation is
suppressed. It would be very interesting to see what would happen
in the case that the metric is also involved. This is very
difficult to do and we have not succeeded in doing this in the
case that all fields are involved.

In this section, we analyze a simple case: three scalar fields and
four U(1) fields are suppressed. This perturbation is in the
direction of $\delta M$$\not=$0, $\delta Q_{A}$=$0$ for all equal
charges. From (5) and (8) we can see that entropy is decreasing
in this perturbation. Varying (1) yields the equations of motion
\bea \label{eq14a} &&
\nabla_{a}\nabla^{a}\phi_{i}+\frac{2}{L^2}\sinh
\phi_{i}-2\sum_{A=1}^4\alpha^{i}_{A}e^{\alpha^{j}_{A}\phi_{j}}(F^{(A)}_{ab})^2=0
\qquad\qquad\qquad\qquad\qquad\qquad\mbox\,{(14.a)}
\nonumber\\
\label{eq14b}
&&\partial_{a}(\sqrt{g}e^{\alpha^{i}_{A}\phi_{i}}(F^{(A)\,ab}))=0
\qquad\qquad\qquad\qquad\qquad\qquad\qquad\qquad\qquad\qquad\qquad\:\,\mbox{(14.b)}
\nonumber\\
\label{eq14c} &&R_{ab}=-\sum_{i=1}^{3} \left\{ \frac{1}{L^2}\cosh
\phi_{i} g_{ab}+\frac{1}{2} \partial_{a} \phi_{i} \partial_{b}
\phi_{i}\right\}+4\sum_{A=1}^{4}\left\{F^{(A)}_{ac}F^{(A)\,c}_{\,\,\,\,\,\,\,\,\,b}-
\frac{1}{4}g_{ab}(F^{(A)})^2\right\}.
\nonumber\\
&&\qquad\qquad\qquad\qquad\qquad\qquad\qquad\qquad\qquad\qquad\qquad
\qquad\qquad\qquad\qquad\qquad\:\:\:\:\,\mbox{(14.c)} \nonumber
\eea We expect  a relevant perturbation is
\setcounter{equation}{14} \bea \label{eq.15}
\delta \phi_{i}&=&0\nonumber\\
\delta F^{(A)}_{ab}&=&0\\
\delta g_{ab}&=&\gamma_{ab}.\nonumber \eea We need to check if
above ansatz is consistent with equations of motion and it is so
if \be \label{eq.16} \gamma=0
\,\,\,\,\,\,\,\,\,\gamma^{t}_{\,\,t}+\gamma^{r}_{\,\, r}=0\ee
where $\gamma=\gamma^{a}_{\,\,a}=g^{ab}\gamma_{ba}$. The linear
perturbation equations from (14.a) and (14.b) are automatically
satisfied for all equal charges . Now we need to check if (15) is
consistent with (14.c). The linear perturbation from (14.c) is

 \bea
&&-\frac{1}{2}\nabla_{a}\nabla_{c}\gamma-\frac{1}{2}\nabla^{b}\nabla_{b}\gamma_{ac}+\nabla^{b}\nabla_{(c}\gamma_{a)b}\nonumber\\
&=&\Lambda
\gamma_{ac}+16\left(-F_{a}^{\,\,b}F_{c}^{\,\,d}\gamma_{bd}-\frac{1}{4}\gamma_{ac}F^2+\frac{1}{2}g_{ac}F^{bd}F^{e}_{\,\,d}\gamma_{be}\right)
\eea where we use the totally symmetric notation for $( \,\,)$,
$\Lambda=-\frac{3}{L^2}$ and $F^2=F^{ab}F_{ab}$. Four U(1) fields
become the same. It is a well-known fact from electromagnetism
that if there is a source term, a simple gauge choice is not easy.
However, if we take the trace of (17), it becomes

\be -\frac{1}{2}\nabla_{a}\nabla^{a}\gamma=\Lambda
\gamma-16\gamma F^2. \ee We used the condition,
$\gamma^{t}_{\,\,t}=-\gamma^{r}_{\,\,r} $ for this. Because (18)
is a homogeneous equation for $\gamma$, we can choose a transverse
traceless gauge whereby
\bea \nabla^{a}\gamma_{ab}&=&0 \nonumber\\
 \gamma&=&0.  \eea See \cite{Wald} for a detail about this gauge
choice. It should be noted that this gauge choice is only
possible with our ansatz that scalar fiends and U(1) fields are
not fluctuating. Finally we get the perturbation equation for the
metric from (17) following \cite{Wald}.  \be (\nabla^{b}
\nabla_{b}+2\Lambda-8F^2)\gamma_{ac}-(R_{c}^{\,\,d}\gamma_{ad}+R_{a}^{\,\,d}\gamma_{cd})-(2R^{b\,\,\,\,d}_{\,\,ac}+32F_{a}^{\,\,b}F_{c}^{\,\,d})\gamma_{bd}=0
\ee It is the even wave in the canonical form \cite{ReggeWheeler}
which is relevant to our case:
\be \gamma_{ab}=  \left( \begin{array}{cccc} \tilde{\gamma}_{tt} & \tilde{\gamma}_{tr} & 0 & 0 \\
\tilde{\gamma}_{rt} & \tilde{\gamma}_{rr} & 0 & 0 \\ 0 & 0 & r^2k & 0 \\
0 & 0 & 0 & r^2k \sin^2 \theta \end{array} \right)
e^{iwt}P_{l}(\cos\theta). \ee It can be easily checked that in
this form, $\gamma^{t}_{\,\,t}$ and $-\gamma^{r}_{\,\,r}$ have the
same equation in (20) and this proves that our ansatz
(\ref{eq.15})-(\ref{eq.16}) is completely consistent with
equations of motion. This equality between $\gamma^{t}_{\,\,t}$
and $-\gamma^{r}_{\,\,r}$ is expected from the $\delta M$
perturbation of the metric in (6). The equations for
$\gamma_{tt}$ and $\gamma_{tr}$ are coupled:
 \bea
&&\left\{-\frac{\partial^2_{t}}{f}+f\partial^2_{r}+\left(2\frac{f}{r}-f'\right)\partial_{r}-2\frac{f'}{r}
+\frac{\partial_{\theta}\sin\theta\partial_{\theta}}{r^2\sin\theta}\right\}\gamma_{tt}+2f'\partial_{t}\gamma_{tr}=0\\
&&\left\{-\frac{\partial^2_{t}}{f}+f\partial^2_{r}+\left(2\frac{f}{r}+f'\right)\partial_{r}-\frac{(f')
^2}{f}-f''+\frac{\partial_{\theta}\sin\theta\partial_{\theta}}{r^2\sin\theta}\right\}\gamma_{tr}+
\frac{2f'\partial_{t}}{f^2}\gamma_{tt}=0 \nonumber\eea Here $f$ is
defined in (6) and $f'=\partial_{r}f$. Using the form (21) we can
make the forth order ordinary differential equation for
$\tilde{\gamma}_{tr}$, which is what we are going to study by
numerics.

\subsection{Numerical analysis}
To carry out a numerical study of (22), we can cast the equation
in terms of a dimensionless radial variable $u$, a dimensionless
charge parameter $\chi$, a dimentionless mass parameter $\sigma$,
and a dimensionless frequency $\tilde{w}$ introduced in
\cite{GM1}:

 \be
u=\frac{r}{M^{\frac{1}{3}}L^{\frac{2}{3}}}\,\,\,\,\,
\chi=\frac{Q}{M^{\frac{2}{3}}L^{\frac{1}{3}}}\,\,\,\,\,
\sigma=\left(\frac{L}{M}\right)^{\frac{2}{3}}\,\,\,\,\,
\tilde{w}=\frac{wL^{\frac{4}{3}}}{M^{\frac{1}{3}}} \ee Then we
combine two equations in (22)

\bea &&
\left[
\left\{ \frac{\tilde{w}^2}{\tilde{f}}+\tilde{f}
\partial^{2}_{u}+\left( 2\frac{\tilde{f}}{u}-\tilde{f}'\right)
\partial_{u}-2\frac{\tilde{f}'}{u}-\frac{\sigma
l(l+1)}{u^2}\right\}\frac{\tilde{f}^2}{2\tilde{f}'}\right.
\\
&\times&
\left.
\left\{\frac{\tilde{w}^2}{\tilde{f}}+\tilde{f}\partial^2_{u}+
\left(
2\frac{\tilde{f}}{u}+\tilde{f}'\right)\partial_{u}-
\frac{(\tilde{f}')^2}{\tilde{f}}-\tilde{f}''-\frac{
\sigma
l(l+1)}{u^2}\right\}+2\tilde{f}'\tilde{w}^2\right]\tilde{\gamma}_{tr}=0\nonumber\\
&&\tilde{f}=\sigma-\frac{2}{u}+\frac{\chi^2}{u^2}+u^2.
\nonumber\eea

Now we need to specify the boundary condition for
$\tilde{\gamma}_{rt}$. We want to place a initial data surface
touching the horizon at one end and ending on the boundary of
$AdS$ at the other end. Because $AdS$ does not have a Cauchy
surface \cite{Hawking}, the future domain of dependence of this
initial data lies inside of Cauchy horizon of $AdS$. To define
`small' for the perturbation at the horizon, we can use
non-singular coordinates, Kruskal coordinates
\cite{Vishveshwara}. Dropping the $S^2$ piece in (6) and
introducing  a tortoise coordinate $r_{\ast}$ and Kruskal
coordinates $(T,R)$ according to \bea
\frac{dr_{\ast}}{dr}&=&\frac{1}{f} \nonumber\\
e^{\frac{1}{2}f'(r_{H})(\pm t+r_{\ast})}&=&\pm T+R.\eea The
near-horizon metric is regular \be ds^2 \approx
\frac{4}{e^{f'(r_{H})r_{H}}f'(r_{H})}(-dT^2+dR^2).\ee We can
express the Kruskal components $\gamma'_{ab}$ in terms of the
original components $\gamma_{ab}$ \bea
&&\gamma'_{tt}=\frac{4}{f'(r_{H})(-T^2+R^2)^2}\left[
R^2\gamma_{tt}-2fRT\gamma_{tr}+f^2T^2\gamma_{rr}\right]\nonumber\\
&&\gamma'_{tr}=\frac{4}{f'(r_{H})(-T^2+R^2)^2}\left[
-TR\gamma_{tt}+2f(T^2+R^2)\gamma_{tr}-f^2TR\gamma_{rr}\right]\\
&&\gamma'_{rr}=\frac{4}{f'(r_{H})(-T^2+R^2)^2}\left[
T^2\gamma_{tt}-2fRT\gamma_{tr}+f^2R^2\gamma_{rr}\right]\nonumber
\eea all of which should be finite as $r \rightarrow r_{H}$ on
our initial data surface. To avoid the issue of mode superposition
and a better physical sense in which black holes would form in a
collapse situation, we would require a surface ending on a future
horizon \cite{Whitt}.  When we approach the future horizon from
outside of a black hole region, $R=T+O(r-r_{H})$. This implies
that normalizable wavefunctions $\gamma_{tr}$ must be
$O(r-r_{H})$ as we approach the event horizon. We also want it
falling off like $1/r^2$ near the boundary of $AdS_4$. This
boundary condition is taken from the asymptotic behavior
$\gamma_{tt} \simeq r^3\gamma_{tr}$. Using Maple, we solved (24)
numerically. We did not find any unstable mode. At $\sigma=0$,
thermodynamic stability is lost at $\chi=1$. The smallest
$\tilde{w}^2$ in this case is $\tilde{w}^2=2$. There is no
normalizable wavefunction with negative $\tilde{w}^2$. Negative
mode is found at $\chi=3.7$, which lies in naked singularity
regions and therefore is not relevant. We can conclude that in
the perturbation, $\delta M$$\not=$0, $\delta Q_{A}$=0 there is
no dynamical instability, which makes sense because this
perturbation would decrease entropy if it were unstable.

\section{Discussions and Perspectives}
We have been working on Gubser-Mitra conjecture, which relates
thermodynamics to  dynamics in black holes. When the metric
fluctuations are suppressed at linear order, all perturbations
are dynamically unstable if black holes are thermodynamically
unstable and all of evolutions increase entropy. However, when
only the metric fluctuations are turned on, they are  dynamically
stable. Entropy is not decreasing in this case. This result
strengthens both the motivation of Gubser-Mitra conjecture, which
claims that Lorentzian time evolution should go so as to increase
entropy, and  the validity of their conjecture. The Second Law of
Thermodynamics is not a rigorous law of nature. It is correct in
a certain macroscopic time scale. Therefore it is remarkable if
dynamics obeys this law, even though we are interested in
classical instabilities.

As suggested in \cite{GM2}, entropic arguments can give a good
information not only on the existence of dynamical instabilities
but also on the direction they point. We can have a very
heuristic argument why there is a dynamical instability when the
perturbation $(\delta M, \delta Q_{A})$ increases entropy.
Entropy can be understood as a functional of the fields
describing black hole. Let $X_{i}(t)$ denote metric, scalar and
U(1) fields. If we write down the equation (8) in terms of
$X_{i}(t)$, it becomes

\be
S(X_{i}(t_{0}+dt))=S(X_{i}(t_{0}))+\frac{1}{2}\left(\frac{\delta^{2}S}
{\delta X_{i}\delta
X_{j}}\right)_{t=t_{0}}\frac{dX_{i}}{dt}\frac{dX_{j}}{dt} dt^2
.\ee Here $X_{i}(t_{0})$ is the solution describing our black hole
and the derivative of entropy with respect to $X_{i}$ is a
functional derivative. $\delta X_{i}(t)$ can be understood as a
solution of the linear perturbation equation in this case. If we
assume that $ \left(\frac{\delta^{2}S} {\delta X_{i}\delta
X_{j}}\right)$ gives a positive definite inner product between
fields, we can see that entropy increases if and only if the time
dependence of $X_{i}(t)$ is $e^{iwt}$ with $w^2$ negative. This
argument does not hold for a Schwarzschild black hole and this
might be due to our assumption on the positivity of $
\left(\frac{\delta^{2}S} {\delta X_{i}\delta X_{j}}\right)$. It
will be a very interesting problem to understand this from the
quantum theory of gravitation and it might explain the reason why
we need a non-compact translational symmetry of a black object in
Gubser-Mitra conjecture.

This is also very interesting in mathematical point of view.
Suppose we have a hypersurface defined by $S=S(M,Q_{A})$. Each
pertuabation $(\delta M, \delta Q_{A})$ corresponds to a tangent
vector originated from $p=(M_0,Q_{0A})$ up to a normalization
factor in a tangent space $T_{p}S$ of the hypersurface at $p$
\footnote{We assume entropy is extremum at p, so that $\delta
S=0$ in this tangent space.}. This tangent space can be
understood as the real projective space  $\Re P^4$. The second
derivative part of equation (8) is a homogeneous polynomial of
$(\delta M, \delta Q_{A})$ of degree 2 and the zero locus of this
polynomial gives us an algebraic subvariety of $\Re P^4$. This
variety separates $\Re P^4$ into two parts: $\delta S > 0 $ and
$\delta S < 0$. We can also separate $\Re P^4$ in another way: A
stable region in which a perturbation $(\delta M, \delta Q_{A})$
gives no dynamical evolution and an unstable region in which a
perturbation $(\delta M, \delta Q_{A})$ gives a dynamical
evolution. If our argument in the previous paragraph is correct,
this means that both of separations are the same and it will give
a very interesting relation between algebraic equations and
differential equations\footnote{Any perturbation $(\delta M,
\delta Q_{A})=(0,\neq 0)$ increases entropy when $\chi>1$.
Therefore we can have linear perturbation equations involving the
metric  which may be dynamically unstable}.

It was argued that a unstable black string settles down to a new
static black string solution  which is not translationally
invariant along the string and can be viewed as a local entropy
maximum but not a global one \cite{Horowitz}. If the final stage
of the evolution is a local entropy maximum, we can not say that
evolutions from different perturbatoins end up with the same
final solution. In our case, we have 3 eigenvectors of Hessian
with the same positive eigenvalues for all equal charge:
$(a=1,b=-1), (a=-1,b=1)$ and $(a=-1,b=-1)$ in (9). Considering
the sign of these vectors, there are six most unstable
perturbation vectors in the tangent space. It would be very
interesting to see that what will be the final solutions for
these perturbations. Finally it is an open question to see if
evolutions from perturbations around the eigenvectors would
result in the same final solutions in which the evolutions from
the  nearby eigenvectors result. We leave these questions for
future work.

\section{Acknowledgments}

We would like to thank V. Balasubramanian, F. A. Brito, M. Cvetic
and M. Strassler for useful discussions. We would like to express
special thanks to A. Naqvi for his help with numerical analysis
and to R. M. Wald for his comprehensive book \cite{Wald}. This
work is supported by DOE grant DE-FG02-95ER40893.


\end{document}